\documentclass[a4paper,10pt]{article}
\usepackage[utf8x]{inputenc}
\usepackage{epsfig}
\usepackage{graphicx}
\usepackage{amsmath}
\usepackage{amssymb}
\usepackage{cite}
\usepackage{fancyhdr}
\newcommand{\lsim}{\raisebox{-0.13cm}{~\shortstack{$<$ \\[-0.07cm] $\sim$}}~} 
\newcommand{\gsim}{\raisebox{-0.13cm}{~\shortstack{$>$ \\[-0.07cm] $\sim$}}~} 

\title{Updated constraints from radiative $\Upsilon$ decays\\ on a light CP-odd Higgs}
\author{{\bf Florian Domingo}\footnote{email: domingo@particle.uni-karlsruhe.de}\vspace{2mm}\\ 
\em Institut f\"ur Theoretische Teilchenphysik\\
\em Karlsruher Institut f\"ur Technologie (Universit\"at Karlsruhe)\\
\em D-76128 Karlsruhe, Germany}
\date{}
\begin{document}
\maketitle
\thispagestyle{fancy}
\fancyhead[R]{TTP10-44\\SFB/CPP-10-91}

\begin{abstract}
The possible existence of a light CP-odd Higgs state in many new-physics models could lead to observable 
effects in the bottomonium sector. Experimental bounds on radiative $\Upsilon$ decays through such a 
pseudoscalar state and possible mixings with the $\eta_b^0$ states are reviewed. Combining these two effects, 
we set constraints on the properties of the CP-odd Higgs in the limit of small photon energy of 
$BR(\Upsilon\to\gamma\,\tau^+\tau^-)$, that is on the pseudoscalar mass-range $\sim8-10$~GeV.
\end{abstract}

\section{Introduction}
The bottomonium sector is well-known as a possibly sensitive probe of light CP-odd Higgs states $A^0$ \cite{
Drees:1989du,SanchisLozano:2002pm,McElrath:2005bp,Sanchis-Lozano:2006gx,Dermisek:2006py,Fullana:2007uq,
Hodgkinson:2008ei}. Such pseudoscalar states have indeed vanishing $V-V-A^0$ ($V=Z,W$) couplings, which allows 
them to circumvent most of the direct collider constraints (contrary to CP-even states). Several new-physics 
models provide a natural framework to embed these particles: MSSM with CP-violating Higgs sector \cite{
Carena:2002bb,Lee:2007ai}, $U(1)$-extensions of the MSSM \cite{Han:2004yd}, Little-Higgs models, 
(non-supersymmetric) Two-Higgs Doublet models \cite{Kraml:2006ga}, etc.\ 

As an illustrative case, we consider the example of the NMSSM (see \cite{Ellwanger:2009dp} for a review), 
where a light CP-odd Higgs can emerge ({\em e.g.}) naturally from an approximate and spontaneously broken R or 
Peccei-Quinn symmetry. A phenomenological possibility in this model is the so-called ``ideal Higgs scenario'', 
where the lightest CP-even Higgs $h_1$ decays unconventionally into a pair of light CP-odd Higgs states with masses 
below the $B-\bar{B}$ threshold \cite{dobr,hsearch1,dg1,Dermisek:2005gg,dg2,hsearch2,Dermisek:2007yt,hsearch3} 
so as to avoid LEP bounds on $e^+e^-\to Z+b\bar{b}$ (since $A^0\to B\bar{B}$ is kinematically forbidden). This 
mechanism allows for comparatively light CP-even Higgs states $\gsim 86$~GeV and can also lead to a successful
interpretation of the $2.3\,\sigma$ excess observed in $e^+e^-\to Z+b\bar{b}$ for $M_{b\bar{b}}\sim98$~GeV 
\cite{Dermisek:2005gg}. The recent analysis of $e^+e^-\to Z+4\tau$ in {\sc aleph} data \cite{ALEPH:2010aw} 
constrains this scenario further but, according to \cite{Dermisek:2010mg}, these bounds might also be 
circumvented by allowing for increased $A^0\to c\bar{c}$ decays; it is otherwise possible to simply increase 
the singlet component of the lightest CP-even state, which has reduced couplings to the SM-sector, {\em e.g.} 
to the $Z$ boson. Note also that the presence of a light CP-odd state in the NMSSM spectrum is not necessarily 
associated with this ``ideal Higgs scenario'' and may as well accompany a heavier CP-even ($h_1$) state (up to 
$m_{h_1}\sim140$~GeV in the NMSSM). Investigating the existence of such an $A^0$ in low-energy signals is thus 
a necessary probe, complementary to Higgs searches at high energy. In this context, the bottomonium sector is 
of particular interest since light CP-odd Higgs states could be produced in $\Upsilon$ decays or mix with 
$\eta_b$-states, provided their coupling to $b$-quarks is sufficiently large \cite{Upsilon,Domingo:2009tb}.

Radiative $\Upsilon$ decays through a CP-odd Higgs state are constrained by several experimental limits, the 
most recent ones originating from $\Upsilon(1S)\to\gamma A^0\,,\ A^0\to l^+l^-$ searches at {\sc Cleo iii} 
\cite{CLEO:2008hs} and $\Upsilon(3S)\to\gamma A^0\,,\ A^0\to l^+l^-$ searches at {\sc BaBar} \cite{
Aubert:2009cka} ($l=\mu,\ \tau$). These experimental limits on a light NMSSM CP-odd Higgs have already been 
studied, in \cite{Dermisek:2006py,Upsilon} for the {\sc Cleo iii} bounds and \cite{Dermisek:2010mg} for the 
{\sc BaBar} bounds, and were shown to constrain most of the significantly coupled region $X_d\gsim0.5-1$ 
($X_d$ being the reduced coupling of the $A^0$ to down-type quarks). 

Such contraints should however be considered with caution in the region $m_A\gsim8$~GeV, where 
$BR(\Upsilon\to\gamma A^0)$ is not well controlled theoretically and could suffer from large corrections. 
Moreover, possibly relevant mixing effects with $\eta_b$ (or $\chi_0$) states could complicate the situation 
further \cite{Drees:1989du,Fullana:2007uq,Domingo:2009tb}. Such limitations leave all the mass region 
$m_A\sim8-10.5$~GeV essentially unconstrained (or unreliably constrained) by these previous analyses and, in 
particular, the higher reach in mass of the {\sc BaBar} bounds ($m_{\tau\tau}\leq10.10$~GeV) would seem 
impossible to exploit.

Alternative searches for a light CP-odd Higgs signal through a breakdown of lepton universality in inclusive 
leptonic $\Upsilon$ decays have likewise failed so far \cite{Guido:2009zz}. Yet, for the same reasons as above, 
few conclusions can be drawn on the mass range $m_A\gsim8$~GeV.

The mixing between the $A^0$ and $\eta_b$ states was also studied and, in particular, it was suggested in 
\cite{Domingo:2009tb} that this effect be responsible for a displacement of the observed $\eta_b(1S)$ mass, 
which (depending on the QCD-based model) is generically lower than what was expected for the hyperfine \
splitting $m_{\Upsilon(1S)}-m_{\eta_b(1S)}$ \cite{Brambilla,eta_b:pqcd}.

The purpose of this paper consists in investigating the impact of the mixing effect on radiative $\Upsilon$ 
decays with small photon energy, that is in the CP-odd mass-range $m_A\sim8-10$~GeV. We show that even when 
the direct decay $\Upsilon(1S/3S)\to\gamma A^0$, which had been considered in \cite{Dermisek:2006py,Upsilon,
Dermisek:2010mg}, is neglected, the additional and indirect contributions $\Upsilon(1S/3S)\to\gamma 
(\eta_b^0\rightsquigarrow A^0)$ resulting from the mixing effect, lead to severe limits on $X_d$ when compared 
to the experimental bounds from {\sc Cleo} and {\sc BaBar}. Given the large uncertainty on hadronic parameters 
and the limited validity of our approximations, such limits should however be considered cautiously and from 
a qualitative, rather than quantitative, viewpoint. In the following section, we summarize the status of 
$BR(\Upsilon\to\gamma A^0(\to l^+l^-))$ in view of the {\sc Cleo} and {\sc BaBar} limits, in the approximation 
of a pure $A^0$ Higgs state. We will also present briefly the mixing effect between the $A^0$ and the $\eta_b$ 
states. In the third section, the consequences for $BR(\Upsilon\to\gamma\,\tau^+\tau^-)$ will be analysed in 
the limit where $m_A\sim m_{\Upsilon}$, under the assumption of negligible direct decay $\Upsilon\to\gamma A^0$. 
The numerical bounds on $X_d$ will be discussed in the last section before a short conclusion.

In the following, we will assume the existence of a NMSSM-like CP-odd Higgs state with mass $m_A\lsim 10.5$ 
GeV and coupling to down-type quarks and leptons of the form $\frac{m_fX_d}{\sqrt{2}\mbox{\em\small v}}$
(where $m_f$ is the fermion mass and $\mbox{\em v}=(2\sqrt{2}G_F)^{-1/2}$ is the electroweak vacuum expectation 
value), while the coupling to up-type quarks is given by $\frac{m_fX_d}{\sqrt{2}\mbox{\em\small v}\,
\tan^2\beta}$. In the NMSSM, $X_d=\cos\theta_A\,\tan\beta$, where $\cos\theta_A$ quantifies the amount of 
doublet-component in $A^0$ while $\tan\beta$ is the usual ratio of the doublet vacuum expectation values. 
Significant corrections to the Yukawa couplings are known to develop at large $\tan\beta$, due to loops of 
supersymmetric particles (see {\em e.g. } \cite{Gorbahn:2009pp}). They would alter the relation 
$X_d=\cos\theta_A\,\tan\beta$ somewhat. Such effects will be neglected here, but we will see that constraints 
from the bottomonium sector can become relevant already at moderate $\tan\beta$. 

Moreover $\tan\beta>1$, so that the decay $A^0\to\tau^+\tau^-$ is dominant (if kinematically allowed). In 
fact, the dominant branching ratios are essentially independent from $\tan\beta$, once $\tan\beta\gsim2-3$
\cite{Dermisek:2010mg}. The results that we present in the following will assume the branching ratios at 
$\tan\beta=5$ (about $0.9$ to $0.75$ for $BR(A^0\to\tau^+\tau^-)$ in the range $m_A=8-10$~GeV; however, 
$BR(A^0\to\tau^+\tau^-)\sim1$ would have been a qualitatively acceptable approximation). Note that for 
$\tan\beta\lsim2$, the coupling of the $A^0$ to $b$-quarks is reduced and would generate little effect on the 
bottomonium sector (except for $\cos\theta_A\sim1$). In the general case, the full $(m_A,\tan\beta)$ 
dependence can be retained. The branching ratio $BR(A^0\to\tau^+\tau^-)$ (also $BR(A^0\to\mu^+\mu^-)$) that we 
use is obtained with the public code NMSSMTools \cite{NMSSMTools}. It was noted that the corresponding value 
can be slightly different from the output of Hdecay \cite{Dermisek:2010mg}. However, such minor effects will 
have little impact on the discussion that follows. Note also that the very-light mass region $m_A\lsim5$~GeV 
is already essentially excluded by low-energy constraints \cite{Andreas:2010ms}, such as $B_s\to\mu^+\mu^-$ or 
$B\to X_s\mu^+\mu^-$ \cite{BNMSSM}. 

Beyond the NMSSM, this analysis should be essentially valid for any model with a light CP-odd state and based 
on a Two-Higgs Doublet model of type II (or even of type I), provided $X_d$ is suitably chosen.

\section{Direct bounds on $BR(\Upsilon\to\gamma A^0)$ and Mixing $A^0-\eta_b$}\label{BRmix}
In first approximation, Upsilon decays through a CP-odd Higgs are described by the so-called Wilczek formula
\cite{Wilczek:1977pj}: 
\begin{equation}\label{eq:wilczek}
\frac{BR(\Upsilon(nS) \to \gamma A^0)}{BR(\Upsilon(nS) \to \mu^+\mu^-)} =
\frac{G_F m_b^2 X_d^2}
{\sqrt{2}\pi\alpha}\left(1-\frac{m_{A}^2}{m_{\Upsilon(nS)}^2}\right)
\times  F 
\end{equation}

This formula was first established in the non-relativistic approximation and the hard-photon limit, {\em i.e.} 
$m_A\ll m_{\Upsilon}$. Several corrections, due to relativistic, QCD and bound-state effects, were then 
included within the factor $F$ (see \cite{guide} for a summary). Yet our control over this factor fails for 
$m_A\gsim 8$~GeV, in the limit of low-energy photons, where large corrections from {\em e.g.} bound-state or 
relativistic effects are expected. Within the known approximate computations, the correction factor $F$ 
vanishes for $m_A\gsim 8.8$~GeV (depending on the $b$-mass), hence leaving this mass range apparently 
unconstrained \cite{Upsilon}. One can indeed expect the $A^0$ contribution to radiative $\Upsilon$ decays to 
vanish in this limit, since, for large wavelengths, the photon simply probes an overall neutral state. Yet, 
the cancellation for $m_A\gsim8.8$~GeV overestimates probably this feature. In fact, we will show in the 
following sections that the mixing effect, for $m_A\sim m_{\eta_b^0}$, leads to sizable contributions, which 
are not captured by eq. (\ref{eq:wilczek}). 

\begin{figure}[t]
 \begin{center}
\includegraphics[width=10cm]{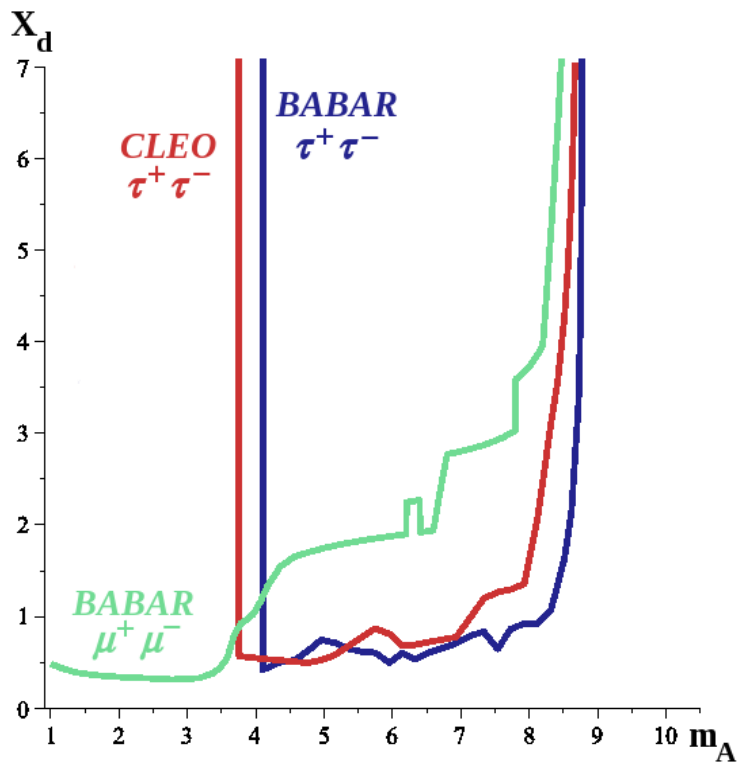}
 \end{center}
\caption{Constraints from $BR(\Upsilon(nS) \to \gamma (A^0 \to \tau^+\tau^-/\mu^+\mu^-))$ on the mass / coupling 
plane of the $A^0$ state. The curves correspond to upper bounds on $X_d$ resulting from the experimental limits. 
The red, dark- and light-blue curves correspond, respectively, to {\sc Cleo} limits on $\Upsilon(1S)\to\gamma
(A^0\to\tau^+\tau^-)$, {\sc BaBar} limits on $\Upsilon(3S)\to\gamma(A^0\to\tau^+\tau^-)$ and {\sc BaBar} limits 
on $\Upsilon(3S)\to\gamma(A^0\to\mu^+\mu^-)$. The correction factor $F$ of eq. (\ref{eq:wilczek}) was assumed 
to be that shown in \cite{Upsilon} and vanishes for $m_A\geq8.8$~GeV.}
\label{fig:botomoniumlim}
\end{figure}

\begin{figure}[t]
 \begin{center}
\includegraphics[width=8.cm]{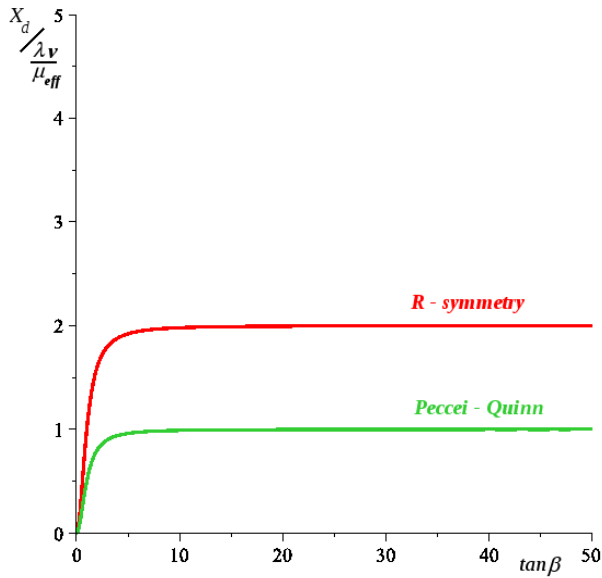}
 \end{center}
\caption{Reduced coupling of the light CP-odd Higgs, $X_d\equiv\cos\theta_A\tan\beta$, as a function of 
$\tan\beta$, in units of $\frac{\lambda\mbox{\em v}}{\mu_{\mbox{\tiny eff}}}$, for the R (red curve) and 
PQ (green) limits. $\lambda$ and $\mu_{\mbox{\tiny eff}}$ are NMSSM parameters \cite{Ellwanger:2009dp}, with 
$\mbox{\em v}\sim175$~GeV, $\lambda\leq0.8$ and $\mu_{\mbox{\tiny eff}}\geq100$~GeV. We thus obtain that $X_d$ 
is generically small in these limits.}
\label{fig:XdRPQ}
\end{figure}

Comparing eq. (\ref{eq:wilczek}) with the experimental limits from {\sc Cleo} \cite{CLEO:2008hs} and 
{\sc BaBar} \cite{Aubert:2009cka} on $BR(\Upsilon(1S/3S)\to\gamma (A^0\to\mu^+\mu^-/\tau^+\tau^-))$, one 
obtains upper bounds on $X_d$ as a function of $m_A$ \cite{Dermisek:2006py,Upsilon,Dermisek:2010mg}, as shown 
in fig. \ref{fig:botomoniumlim}. For the mass region beneath $m_A\sim8.8$~GeV, only moderate to small values 
of $X_d\lsim0.5$ are allowed. Note that this region of weak coupling is however that which is usually favoured 
from the theoretical point of view: the R and Peccei-Quinn symmetry limits of the NMSSM would naturally 
predict small $X_d$ (see fig. \ref{fig:XdRPQ}). On the other hand, such weakly-coupled $A^0$ will prove 
difficult to observe experimentally.

As we discussed in the introduction, the mass range $m_A\sim9-10.5$~GeV is (due to the smallness of the factor 
$F$ in eq. (\ref{eq:wilczek}), within the known computations) free of the latter constraints, so that large 
values of $X_d$ seem {\em a priori} allowed. In this range however, one expects to find the $\eta_b(nS)$, 
$n=1,2,3$, states. Such states carry the same quantum numbers as the CP-odd Higgs, so that a mixing 
$A^0-\eta_b$ develops. This scenario was studied in \cite{Drees:1989du,Fullana:2007uq,Domingo:2009tb}. 
We shall use the notations of \cite{Domingo:2009tb}. The mixing can be described by an effective mass-matrix 
for the $\eta_b-A^0$ states:
\begin{equation}\label{eq:massmatr}
{\cal M}^2=\begin{pmatrix}
     m_{\eta_b^0(1S)}^2 & 0 & 0 & \delta m_1^2\\
     0 & m_{\eta_b^0(2S)}^2 & 0 &\delta m_2^2\\
     0 & 0 & m_{\eta_b^0(3S)}^2 & \delta m_3^2\\
     \delta m_1^2 & \delta m_2^2 & \delta m_3^2 & m_{A}^2
\end{pmatrix}
\end{equation}
The diagonal elements $m_{\eta_b^0(nS)}^2$, $n=1,2,3$, correspond to the masses of the pure QCD $b\bar{b}$ 
states which can be estimated through the hyperfine-splitting $m^2_{\Upsilon(nS)}-m^2_{\eta_b^0(nS)}$ 
\cite{eta_b:pqcd}:
\begin{equation}
m_{\eta_b^0(1S)} \simeq 9418 \pm 13\ \mathrm{MeV}\ \ ,\ \ m_{\eta_b^0(2S)} \simeq 10002~\mbox{MeV}
\ \ ,\ \ m_{\eta_b^0(3S)} \simeq 10343~\mbox{MeV}\label{etabpqcd}
\end{equation}
The quoted value for $m_{\eta_b^0(1S)}$ is the perturbative QCD (pQCD) prediction. Slightly different 
predictions can be obtained in other approaches \cite{Brambilla}. The uncertainty on the heavier 
$\eta_b^0$ masses will be neglected. The off-diagonal terms $\delta m_n^2$ can be computed in a 
non-relativistic quark-potential model \cite{Drees:1989du,Fullana:2007uq,Upsilon}:
\begin{equation}\label{dmest}
\delta m_n^2\ =\  
\biggl(\frac{3m_{\eta_{b}(nS)}^3}{8 \pi v^2}\biggr)^{1/2}
|R_{\eta_b(nS)}(0)|
\times X_d\;\ \ \ \Rightarrow\ \ \ \left\{\begin{array}{ll}
\delta m_1^2&\simeq(0.14\pm 10\%)\ \mathrm{GeV}^2\times X_d,\\
 \delta m_2^2&\simeq(0.11\pm 10\%)\ \mathrm{GeV}^2\times X_d, \\
 \delta m_3^2&\simeq(0.10\pm 10\%)\ \mathrm{GeV}^2\times X_d\end{array}\right.
\end{equation}
where the $\eta_b^0(nS)$ wave-functions $R_{\eta_b(nS)}$ were estimated through those of the $\Upsilon(nS)$ 
states, which can be extracted, in turn, from the leptonic decays $\Upsilon(nS)\to l^+l^-$ \cite{PDG}. The 
other off-diagonal entries vanish as a result of the orthogonality between $\eta_b^0$ states. The physical 
mass-states are denoted as:
\begin{equation}\label{massst}
 \eta_i=P_{i1}\,\eta_b^0(1S)+P_{i2}\,\eta_b^0(2S)+P_{i3}\,\eta_b^0(3S)+P_{i4}\,A^0
\end{equation}
\begin{figure}[t]
 \begin{center}
\includegraphics[width=10cm]{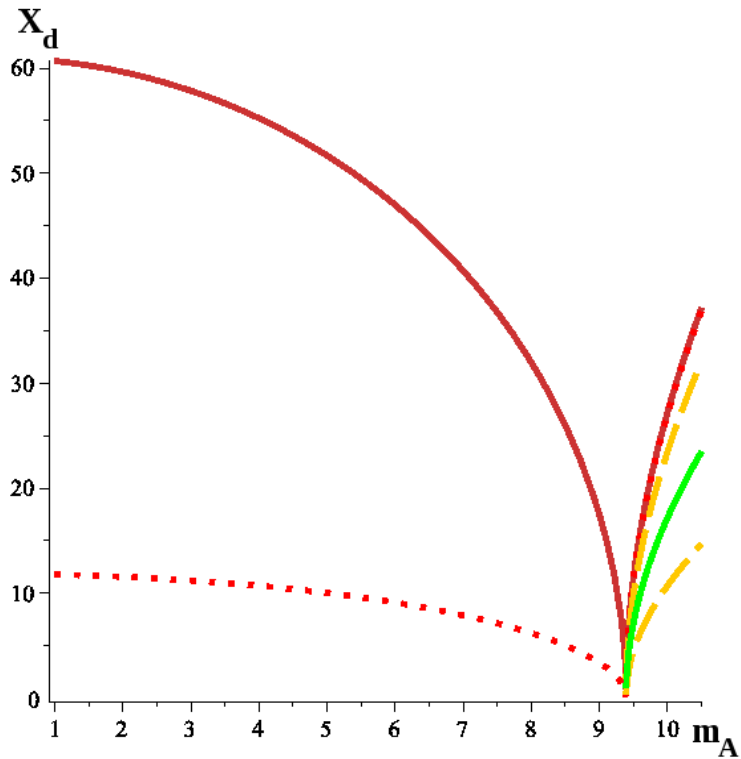}
 \end{center}
\caption{Relations and constraints in the plane $(m_A,X_d)$ originating from the splitting effect in the pseudoscalar 
mass matrix and the observed $\eta_b$ mass. The red curves represent upper bounds on $X_d$: the full line 
corresponds to the conservative limits $-30~\mbox{MeV}\leq m_{\eta_b^0(1S)}-m_{\mbox{\tiny obs}}\leq 
54~\mbox{MeV}$; the dotted one corresponds to the pQCD $2\,\sigma$ range. In the domain between the orange 
dashed curves, the splitting between the observed $\eta_b$ mass and the pQCD prediction is generated by the 
mixing effect within $1\,\sigma$. The green curve is the ``favoured'' line obtained with the central value of 
the pQCD prediction \cite{Domingo:2009tb}.}
\label{fig:mixbound}
\end{figure}

The only experimental information at our disposal concerning the mass matrix of eq. (\ref{eq:massmatr}) 
originates from the observation at {\sc BaBar}, in $\Upsilon(3S)$ \cite{BABAR:2008vj} and $\Upsilon(2S)$ 
\cite{BABAR:2009pz} decays, of a state with mass $m_{\mbox{\tiny obs}}=9390.9\pm3.1$~MeV \cite{BABAR:2009pz} 
which was interpreted as the $\eta_b^0(1S)$. It is remarkable that this value for the mass is in slight 
tension with the pQCD prediction (eq. (\ref{etabpqcd})). Within most of the QCD models, one expects a slightly 
smaller hyperfine splitting \cite{Brambilla} (although possibly within $1\,\sigma$). As a consequence, the 
mixing effect with the $A^0$ was suggested as a possible interpretation of a displaced $\eta_b^0(1S)$ mass 
\cite{Domingo:2009tb}. The requirement that $m_{\mbox{\tiny obs}}^2$ be an eigenvalue of the mass matrix 
(\ref{eq:massmatr}) leads to the relation:
\begin{equation}
m_{A}^2 = m_{\mbox{\tiny obs}}^2 
+ \frac{\delta m_1^4}{m_{\eta_b^0(1S)}^2 - m_{\mbox{\tiny obs}}^2}
+ \frac{\delta m_2^4}{m_{\eta_b^0(2S)}^2 - m_{\mbox{\tiny obs}}^2}
+ \frac{\delta m_3^4}{m_{\eta_b^0(3S)}^2 - m_{\mbox{\tiny obs}}^2}
\label{eq:ma}
\end{equation}
For given QCD-predicted $\eta_b$ masses (see eq. (\ref{etabpqcd})) and the mixing elements of eq. 
(\ref{dmest}), eq. (\ref{eq:ma}) determines $m_A$ in terms of $X_d$. Varying the input from eq. 
(\ref{etabpqcd}) and (\ref{dmest}) within the error bars, this relation results in bounds on the 
plane $(m_A,X_d)$ \cite{Upsilon}. These limits originate simply from the observation that, in the presence of 
a mixing, a mass-splitting is generated so that the observed mass cannot coincide with a diagonal entry of the 
mass matrix (to which it couples). Assuming the pQCD result to be valid, one can define a favoured region in 
the plane $(m_A,X_d)$ where the observed mass can be reproduced within $1\,\sigma$. This is illustrated in 
fig. \ref{fig:mixbound}. The $2\,\sigma$ limits are also shown. However, given the larger number of 
predictions in alternative QCD-models, we retain, in what follows, the bounds resulting from the more 
conservative range $-30~\mbox{MeV}\leq m_{\eta_b^0(1S)}-m_{\mbox{\tiny obs}}\leq 54~\mbox{MeV}$.

\section{Radiative $\Upsilon$ decays in the mixing scenario}\label{sec:formal}
\begin{figure}[t]
 \begin{center}
\includegraphics[width=12.cm]{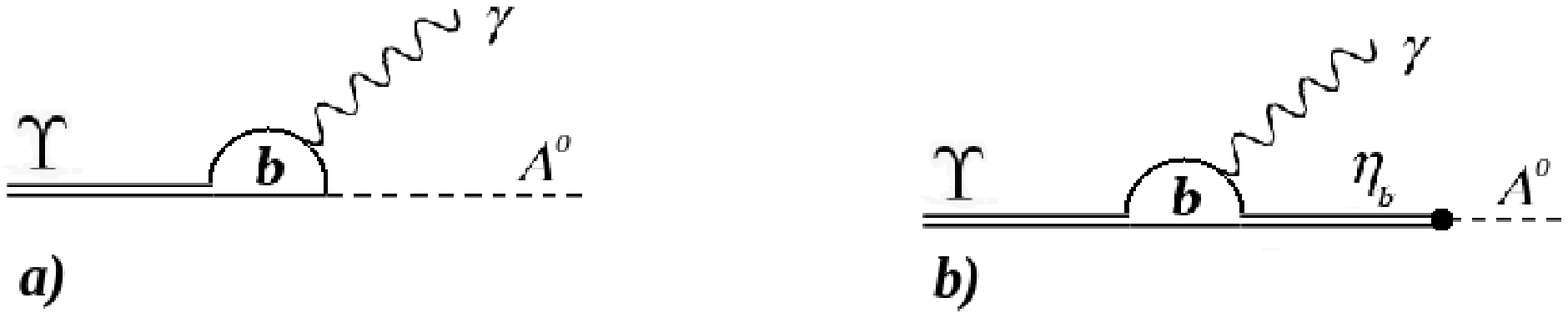}
 \end{center}
\caption{Contributions to the radiative Upsilon decay to a Higgs pseudoscalar.\newline
a) Direct decay, corresponding to the description of eq. (\ref{eq:wilczek}).\newline
b) Additional mixing contribution due to the $A^0-\eta_b$ mixing.}
\label{fig:bottometab}
\end{figure}

$BR(\Upsilon(nS) \to \gamma A^0)$, as we presented it in the previous section, corresponds to the diagram of 
fig. \ref{fig:bottometab}a. In the presence of a $A^0-\eta_b$ mixing, however, a second contribution, as in 
fig. \ref{fig:bottometab}b, arises. This contribution was already described in \cite{Fullana:2007uq} under the 
denomination ``resonant decay''. Assuming that the contribution of fig. \ref{fig:bottometab}a vanishes for 
small $A^0$-$\Upsilon$ mass differences, as we discussed in section \ref{BRmix}, the diagram of fig. 
\ref{fig:bottometab}b is then dominant and could lead to a significant effect. In this section, we will 
explicitely neglect this diagram of fig. \ref{fig:bottometab}a when $m_A\gsim 8$~GeV and focus exclusively on 
the contribution b due to mixing.

Let us first consider the transition between pure bottomonium states. In the non-relativistic approach to 
quarkonium bound states, $BR(\Upsilon(nS)\to\gamma \eta_b^0(jS))$ can be written as \cite{Godfrey:2001eb}:
\begin{equation}\label{eq:BRetab}
 BR(\Upsilon(nS)\to\gamma \eta_b^0(jS))=\frac{4\ \alpha\, e_b^2}{3\,m_b^2\,\Gamma_{\Upsilon(nS)}}|I_{nj}|^2
\,k^3\,\cdot\Theta(k)
\end{equation}
where $\alpha=(137.036)^{-1}$ is the fine-structure constant, $e_b=-1/3$, the $b$-quark charge, $k$, the 
photon energy and $\Gamma_{\Upsilon(nS)}$, the $\Upsilon(nS)$ total width ($\Theta$ is the Heaviside 
distribution). We shall use $\Gamma_{\Upsilon(1S)}=(54.02\pm 1.25)$~keV and $\Gamma_{\Upsilon(3S)}=
(20.32\pm1.85)$~keV \cite{PDG}. The transition factor $I_{nj}$ is defined as:
\begin{equation}\label{eq:ifact}
 I_{nj}=\left<\eta_b^0(jS)\right|j_0(kr/2)\left|\Upsilon(nS)\right>
\end{equation}
where $j_0(x)=\frac{\sin x}{x}$. Taking $r_0\sim1.4~\mbox{GeV}^{-1}$ as a typical scale of confinement, we see 
that this function can be safely expanded in the limit of low-energy photons as $j_0(kr/2)\simeq1-(kr/2)^2/6+\ldots$. 
Using that the wave functions for the $\eta_b^0$ and $\Upsilon$ states are almost identical, one obtains that 
in first approximation $I_{nj}\simeq \delta_{nj}$: $nS\to nS$ transitions are {\em a priori} favoured. The 
resulting $k^3$ dependence in eq. (\ref{eq:BRetab}) is then characteristic of point-like interactions of the 
bottomonium states, which could be anticipated for large wavelengths.

Note however that for QCD $b\bar{b}$ states, the photon energies are much smaller for a $nS\to nS$ transition 
than for a $3S\to1S/2S$ transition, resulting in comparable branching ratios. The coefficients $I_{nj}$, 
$n\neq j$, were estimated in QCD (see {\em e.g.} \cite{Godfrey:2001eb}), taking into account relativistic 
corrections, but such predictions range over one to two orders of magnitude. Fortunately, this situation will 
be slightly changed in the case of mixed states (see eq. (\ref{eq:Upsgameta0}) below) and it will be possible 
to focus on the dominant $nS\to nS$ channel. For the time being, let us use that $I_{nj}\simeq \delta_{nj}$ and 
that photon energies are small, and parametrize:
\begin{equation}\label{eq:I}
 I_{nj}\simeq \delta_{nj}-k^2\cdot R_{nj}^2
\end{equation}
So far the $R_{nj}^2$ coefficients are simply unknowns, which might be extracted from the QCD computations of 
the $I_{nj}$. Under the assumption that the $nS\to nS$ transition dominates, it will be possible to neglect the 
$R_{nj}^2$, $n\neq j$.

In fact, for large mixing (that is, for $m_A$ in the vicinity of the $\eta_b^0$ masses), processes involving 
the CP-odd Higgs or the $\eta_b^0$ states are more adequately described in the context of the mixing formalism 
which we have presented in section \ref{BRmix}. In particular, the production of any mass state $\eta_i$, 
$i=1-4$, in radiative $\Upsilon$ decays may be obtained (under the assumption that the contribution of fig. 
\ref{fig:bottometab}a is negligible) by a generalization of eq. (\ref{eq:BRetab}) as:
\begin{equation}\label{eq:Upsgameta0}
 BR(\Upsilon(nS)\to\gamma \eta_i)=\frac{4\ \alpha\, e_b^2}{3\,m_b^2\,\Gamma_{\Upsilon(nS)}}\,k_i^3\cdot
\Theta(k_i)\,\sum_{j,l=1}^3 I_{nj}^*I_{nl}\,P_{ij}P_{il}
\end{equation}
Note that the photon energy is $k_i\equiv\frac{m_{\Upsilon(nS)}^2-m_{\eta_i}^2}{2m_{\Upsilon(nS)}}$ for all the 
terms on the right-hand side of eq. (\ref{eq:Upsgameta0}), including pure $nS\to nS$ transitions as well as 
$nS\to \tilde{n}S$, $n\neq\tilde{n}$, transitions or interference terms. Under such conditions, the $nS\to nS$
transition dominates eq. (\ref{eq:Upsgameta0}) due to its larger transition factor. In the following, we 
will neglect the interference terms in eq. (\ref{eq:Upsgameta0}): such terms cannot be large unless the 
$\eta_i$ has simultaneously large $\eta_b^0(jS)$ and $\eta_b^0(lS)$, $j\neq l$, components, which (almost) never 
happens since the $\eta_b^0$ states are orthogonal; moreover, both transition factors $I_{nj}$, $I_{nl}$, 
$j\neq l$,  would have to be large, which contradicts eq. (\ref{eq:I}) for small photon energy. Eq. 
(\ref{eq:Upsgameta0}) thus reduces to:
\begin{equation}\label{eq:Upsgameta}
BR(\Upsilon(nS)\to\gamma \eta_i)\simeq\sum_{j=1}^3P_{ij}^2\,BR_i(\Upsilon(nS)\to\gamma \eta_b^0(jS))
\end{equation}
where $BR_i(\Upsilon(nS)\to\gamma \eta_b^0(jS))$ is given by eq. (\ref{eq:BRetab}), while the index $i$ reminds 
us however that the relevant mass for the $\eta^0$ state is that of the mass state $\eta_i$. Note that the 
simplification from eq. (\ref{eq:Upsgameta0}) to eq. (\ref{eq:Upsgameta}) will work well due to the dominant 
$nS\to nS$ channel. This also implies, however, that we restrict ourselves to small photon energies ($k_i$), 
so that $I_{nj}\simeq\delta_{nj}$ remains a good approximation: therefore, we will require in the following 
that $k_i\lsim 1$~GeV.

Then, the decay of the mass state $\eta_i$ to a tauonic pair can be assumed to proceed through the $A^0$ 
component \cite{Domingo:2009tb}:
\begin{equation}\label{eq:br1}
BR(\eta_i \to \tau^+\tau^-) = \frac{P_{i4}^2 \,\Gamma_{i}(A^0\to \tau^+\tau^-)}
{\sum_{j=1}^3P_{ij}^2 \Gamma_{\eta_b^0(jS)} + P_{i4}^2 \Gamma_{A^0}}
\end{equation}
where the widths $\Gamma_{\eta_b^0(1,2,3S)}$ can be estimated to $10$, $5$ and $5$~MeV respectively\footnote{
After $\Gamma_{\eta_b^0(nS)}/\Gamma_{\eta_c^0(nS)}\simeq (m_b/m_c)[\alpha_S(m_b)/\alpha_S(m_c)]^5\simeq0.25-
0.75$ \cite{oliver}.}.  $\Gamma_{A^0}$ and $\Gamma_{i}(A^0\to \tau^+\tau^-)$ can be computed. Again, the index 
$i$ indicates that the partial width should be calculated using the mass of the state $\eta_i$ instead of that 
of the diagonal $A^0$. The $BR(\eta_i \to \tau^+\tau^-)$ has already been proposed as a probe for light 
pseudoscalars in \cite{Rashed:2010jp}.

We can eventually write the radiative $\Upsilon$ decay to a tauonic pair as:
\begin{equation}\label{eq:Upsgameta2}
 BR(\Upsilon(nS)\to\gamma\, \tau^+\tau^-)\simeq\sum_{i=1}^4BR(\Upsilon(nS)\to\gamma \eta_i)\cdot BR(\eta_i\to 
\tau^+\tau^-)
\end{equation}
where $BR(\Upsilon(nS)\to\gamma \eta_i)$ and $BR(\eta_i\to \tau^+\tau^-)$ are given respectively by eq. 
(\ref{eq:Upsgameta}) and (\ref{eq:br1}). We stress that, in this analysis, the main contribution to eq. 
(\ref{eq:Upsgameta2}) is associated with the subprocess:
\begin{equation}
 \Upsilon(nS)\to\gamma\ \left(\ \ \eta_b^0(nS)\rightsquigarrow A^0\to\tau^+\tau^-\ \ \right)
\end{equation}
the $nS\to nS$ transition being favoured by its large transition factor while the mixing effect allows it to 
access photon energies comparable to those of the other transitions. 

At this point, neglecting the direct contribution $\Upsilon\to\gamma A^0$ of fig. \ref{fig:bottometab}a was 
the most important assumption. If the factor $F$ of eq. (\ref{eq:wilczek}) is indeed small in the considered 
mass-range ($m_A\sim8-10$~GeV), as its cancellation in known computations seems to indicate, we may infer that 
the amplitude of fig. \ref{fig:bottometab}a remains small, so that our assumption in neglecting the direct 
contribution is justified. In practice, the limits on $X_d$ that we obtain (in section \ref{sec:bounds}) using 
the indirect decay (fig. \ref{fig:bottometab}b) only, are comparable in the mass range $m_A=8-10$~GeV to 
those that eq. (\ref{eq:wilczek}) alone would have given with $F\sim0.5-1$ (for the {\sc BaBar} bounds) or 
$F\sim1$ (for the {\sc Cleo} bounds): therefore, we conclude that a factor $F\lsim0.1$ should be a sufficient 
condition to ensure that the contribution of fig. \ref{fig:bottometab}b captures the dominant effect, hence 
that our approach is valid. Note that, among the neglected terms, the non-interferent contribution given by 
eq. (\ref{eq:wilczek}) could be directly included in eq. (\ref{eq:Upsgameta}) (if $F$ were satisfactorily 
kwown) and, being positive, would only strengthen the bounds that we obtain in section \ref{sec:bounds}. 
However, if the amplitude of fig. \ref{fig:bottometab}a were non-negligible, the interference between the 
diagrams of fig. \ref{fig:bottometab}a and \ref{fig:bottometab}b could be significant: since its sign is not 
controlled, this would spoil the validity of our analysis.

A remark concerning the choice of the diagonal $\eta_b^0(1S)$ mass is necessary. In the presence of the 
mixing structure of eq. (\ref{eq:massmatr}) and knowing one of the eigenvalues $m_{\mbox{\tiny obs}}$, one 
cannot choose $m_{\eta_b^0(1S)}$, $m_A$ and $X_d$ independently (or $m_{\mbox{\tiny obs}}$ will not be 
recovered in the general case). Therefore, we will determine $m_{\eta_b^0(1S)}$ in terms of $m_A$, $X_d$ and 
$m_{\mbox{\tiny obs}}$ through eq. (\ref{eq:ma}). Should $m_{\eta_b^0(1S)}$ reach unrealistic values 
(that is, if the condition $-30~\mbox{MeV}\leq m_{\eta_b^0(1S)}-m_{\mbox{\tiny obs}}\leq 54~\mbox{MeV}$ is 
violated), the corresponding points will be excluded through the mixing contraint (see end of section 
\ref{BRmix} and fig. \ref{fig:mixbound}).

Finally, the advantage of considering the three $\eta_b^0$ states simultaneously lies in a more accurate and 
conservative description of the $\Upsilon(nS)\to\gamma\eta_i$ transition: whereas the $\eta_b^0(j=nS)$ 
component is the most relevant one, the presence of the other $\eta_b^0(j\neq nS)$ components reduces the size 
of the corresponding mixing element, hence alleviates slightly the impact of the experimental bounds.

We thus conclude that, if only through the contribution of fig. \ref{fig:bottometab}b, the experimental bounds 
on $BR(\Upsilon(nS)\to\gamma\,\tau^+\tau^-)$ should translate into constraints on the plane $(m_A,X_d)$, even 
in the limit $m_A\sim9-10$~GeV.

\section{Bounds on the light CP-odd Higgs}\label{sec:bounds}
In this section, we present the consequences of the analysis of section \ref{sec:formal} for the light CP-odd 
Higgs in the limit of small photon energies. Given the large uncertainties on the transition factors $I_{nj}$, one cannot 
expect to obtain strict bounds. Instead, we evaluate eq. (\ref{eq:Upsgameta2}) under several approximations 
and compare the results to the experimental limits.
\begin{figure}[h!]
 \begin{center}
\includegraphics[width=15.5cm]{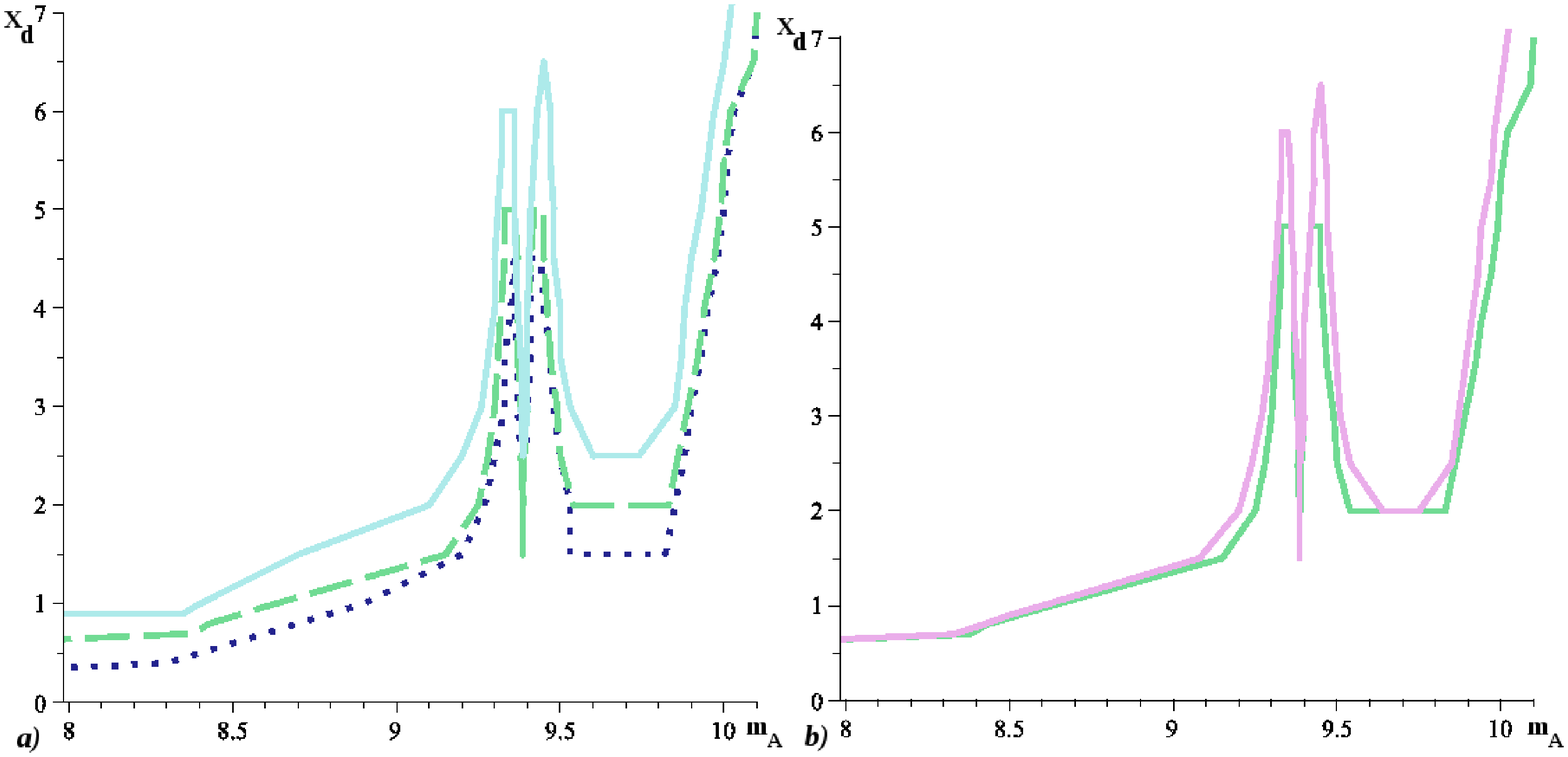}
 \end{center}
\vspace{-.8cm}
\caption{Upper bounds on $X_d$ as a function of $m_A$ due to the mixing effect in $BR(\Upsilon(3S)\to\gamma\, 
\tau^+\tau^-)$ (using {\sc BaBar} data).\newline
a) Influence of the $I_{nj}$ coefficients on the bounds. $\Gamma_{\eta_b^0(1,2,3S)}=10,5,5$~MeV respectively. 
The full light-blue curve assumes $R_{3j}^2=0$ and $|I_{33}|^2=1/2$; the dashed middle-blue curve corresponds 
to $R_{3j}^2=0$ and $|I_{33}|^2=1$; the dotted dark-blue curve assumes the $R_{3j}^2$ as described in the main 
text.\newline
b) Influence of the $\eta_b^0$ widths on the bounds. $R_{3j}^2=0$ and $|I_{33}|^2=1$, with 
$\Gamma_{\eta_b^0(1,2,3S)}=10,5,5$~MeV (middle-blue curve) and $\Gamma_{\eta_b^0(1,2,3S)}=20,10,10$~MeV (pink 
curve).}
\label{fig:mixUps3S}
 \begin{center}
\includegraphics[width=15.5cm]{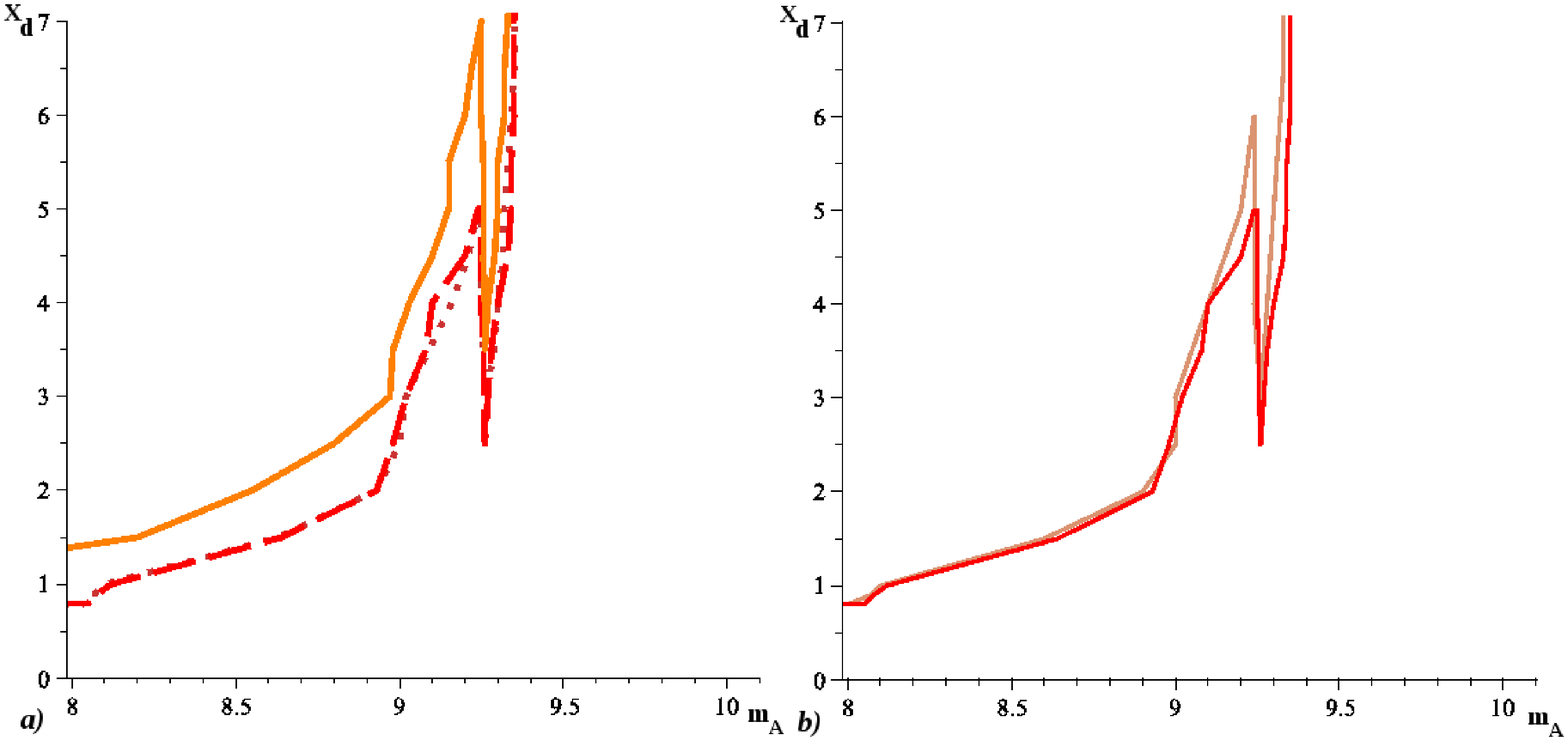}
 \end{center}
\vspace{-.8cm}
\caption{Upper bounds on $X_d$ as a function of $m_A$ due to the mixing effect in $BR(\Upsilon(1S)\to\gamma\, 
\tau^+\tau^-)$ (using {\sc Cleo} data).\newline
a) Influence of the $I_{nj}$ coefficients on the bounds. $\Gamma_{\eta_b^0(1,2,3S)}=10,5,5$~MeV respectively. 
The full orange curve assumes $R_{1j}^2=0$ and $|I_{11}|^2=1/2$; the dashed red curve corresponds to 
$R_{1j}^2=0$ and $|I_{11}|^2=1$; the dotted dark-red curve assumes the $R_{1j}^2$ as described in the main 
text.\newline
b) Influence of the $\eta_b^0$ widths on the bounds. $R_{1j}^2=0$ and $|I_{11}|^2=1$, with 
$\Gamma_{\eta_b^0(1,2,3S)}=10,5,5$~MeV (red curve) and $\Gamma_{\eta_b^0(1,2,3S)}=20,10,10$~MeV (brown curve).}
\label{fig:mixUps1S}
\end{figure}

The {\sc BaBar} limits on $BR(\Upsilon(3S)\to\gamma\,\tau^+\tau^-)$ have the highest reach in mass ($
m_{\tau\tau}\leq10.10$~GeV, limited kinematically by the $\Upsilon(3S)$ mass and the requirement of observable, 
not-too-soft photons), hence are of particular interest. As a first approach, we will consider only the 
$3S\to3S$ transition with the approximation that $|I_{33}|^2\simeq1$, neglecting the coefficients $R_{nj}^2$. 
Then we will continue to neglect the transition factors $I_{nj}$, $n\neq j$, but assume $|I_{33}|^2$ is reduced 
down to $1/2$. Finally, we will use eq. (\ref{eq:I}) with the values $R_{31}^2\sim0.02~\mbox{GeV}^{-2}$, 
$R_{32}^2\sim0.25~\mbox{GeV}^{-2}$ and $R_{33}^2\sim0.02~\mbox{GeV}^{-2}$. This choice of $R_{31}^2$ allows to 
reproduce approximately the measured $BR(\Upsilon(3S)\to\gamma\eta_{\mbox{\tiny obs}})=(4.8\pm1.3)\cdot10^{-4}$ 
\cite{BABAR:2008vj} in the limit where no mixing is assumed. $R_{32}^2$ was chosen so as to have the same 
approximate ratio with $R_{31}^2$ as in \cite{Godfrey:2001eb}. $R_{33}^2$ is simply the order $1$ term in the 
expansion of $j_0(kr_0/2)$, $r_0$ being the typical scale of confinement. The choice of these values is of 
course arguable: the main purpose here consists in checking that the leading effect is captured by the 
$3S\to3S$ transition. The corresponding limits on the $(m_A,X_d)$ plane are shown in fig. \ref{fig:mixUps3S}a. 
We varied all the experimental input within $2\,\sigma$ before extracting the bounds. The $\eta_b^0$ widths 
were however kept at $10$, $5$ and $5$~MeV respectively: their effect is studied separately, in fig. 
\ref{fig:mixUps3S}b. We observe that values of $X_d\gsim 1-2$ are constrained severely and more or less 
equivalently in the three cases we considered and we regard this fact as a sign of robustness of the 
corresponding limits: despite significant variations of the $I_{nj}$ factors in the three cases under consideration,
the excluded region in the $(m_A,X_d)$ plane remains essentially unchanged. The underlying reason for this is 
related to the dominant $3S\to3S$ transition. Fig. \ref{fig:mixUps3S}b leads to a similar observation: 
increased $\eta_b^0$ widths alleviate slightly the bounds as the total widths of the $\eta_i$ state is 
increased (leading to a smaller branching ratio into $\tau^+\tau^-$). Yet despite a doubling of the 
(conjectured) $\eta_b^0$ widths, the bounds on the properties of the $A^0$ remain little affected. The dip 
for $m_A\sim m_{\mbox{\tiny obs}}$ corresponds to unacceptable values of $m_{\eta_b^0(1S)}$. For 
$m_A\lsim 9.3$~GeV, the lightest state (dominantly $A^0$) leads to larger photon energies so that the plotted 
bounds become unreliable. 

Although our approach to the {\sc BaBar} limits should fail for $m_A\lsim 9.3$~GeV, the {\sc Cleo} limits can 
then be applied. The favoured transition is now $1S\to1S$. As before, we consider the cases 
$|I_{11}|^2\simeq1$, $|I_{11}|^2\simeq1/2$ and finally $R_{11}^2\sim R_{33}^2$, $R_{12}^2\sim R_{32}^2$, 
$R_{13}^2\sim R_{31}^2$. The result is shown in fig. \ref{fig:mixUps1S} and constrains $X_d\lsim2-4$ in the 
range $m_A\sim8.4-9.4$~GeV. Again, we observe little variations of the general aspect of the bounds in the 
several considered cases.

\begin{figure}[t]
 \begin{center}
\includegraphics[width=12.cm]{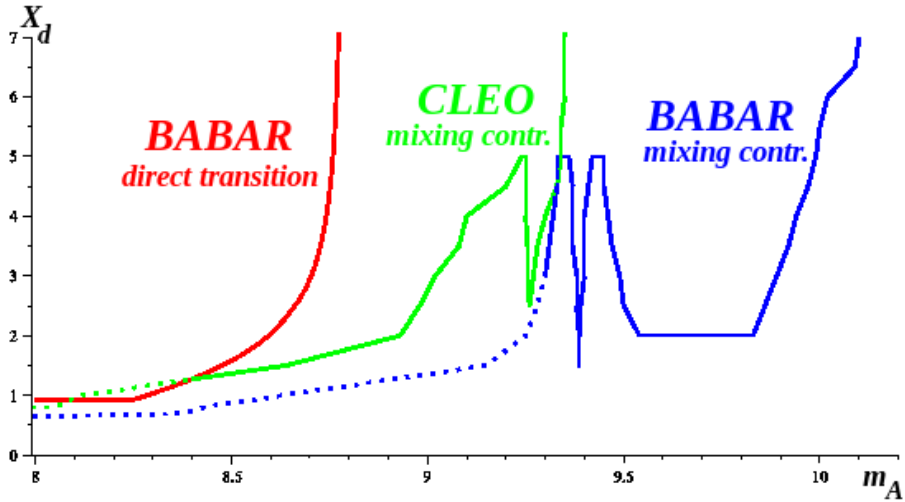}
 \end{center}
\caption{Conclusion: upper bounds on $X_d$ for a pseudoscalar $A^0$ in the mass-range $8-10.1$~GeV.\newline
\hspace{1mm}- in red, from {\sc BaBar}, resulting from the direct $\Upsilon$ decay to a $A^0$ state (fig. 
\ref{fig:bottometab}a): similar to fig. \ref{fig:botomoniumlim};\newline
\hspace{1mm}- in green, from {\sc Cleo}, resulting from the mixed contribution (fig. \ref{fig:bottometab}b); 
the full line corresponds to the approximate validity range, while the dotted curve extends to regions where
the condition on the photon energy $k\leq1$~GeV is not satisfied ({\em i.e.}, where the assumption 
$I_{nn}\sim1$, hence the corresponding bounds, are unreliable);\newline
\hspace{1mm}- in blue, from {\sc BaBar}, resulting from the mixed contribution (fig. \ref{fig:bottometab}b); 
the full line corresponds to the approximate validity range, while the dotted curve extends to regions where
the condition on the photon energy $k\leq1$~GeV is not satisfied.
}
\label{fig:conclim}
\end{figure}

\begin{figure}[t]
 \begin{center}
\includegraphics[width=10.cm]{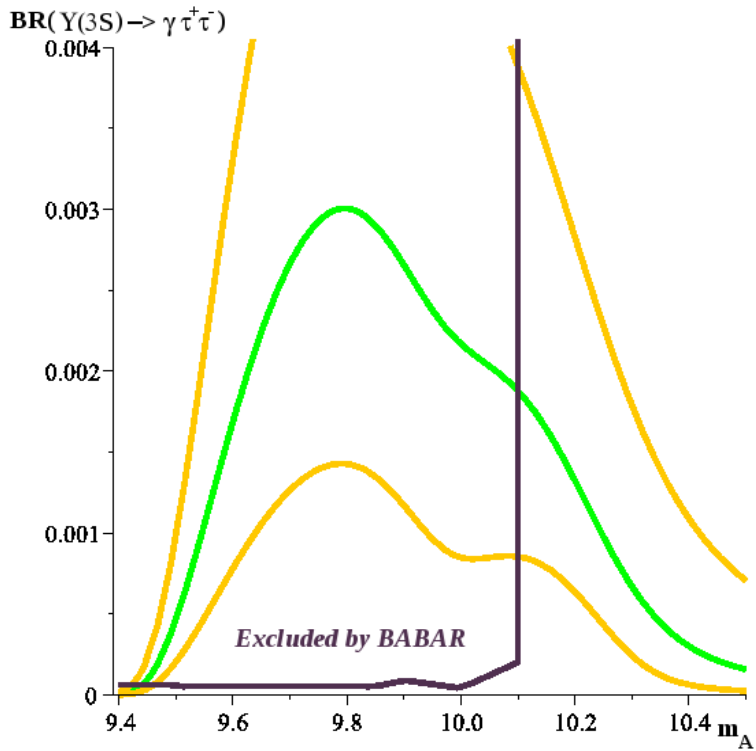}
 \end{center}
\caption{Mixed $\eta_b^0-A^0$ contribution to $BR(\Upsilon(3S)\to\gamma \tau^+\tau^-)$ compared to the 
{\sc BaBar} limits (dark curve) in the ``favoured'' scenario of \cite{Domingo:2009tb}. The result is shown 
along the ``favoured'' line (green curve), and the two $1\,\sigma$ boundaries (orange curves). Except for a 
few $10$~MeV around $m_{\mbox{\tiny obs}}\sim 9.39$~GeV and the large mass region ($m_A\geq10.1$~GeV), 
unconstrained by {\sc BaBar}, the essential part of the mass range in this scenario would lead to excessive 
contributions.
}
\label{fig:mixUps3SBR}
\end{figure}

We conclude this analysis with fig. \ref{fig:conclim} where we collect all the previously discussed limits on 
the $(m_A,X_d)$ plane in their approximate range of validity. (We chose the case $I_{nn}=1$, $I_{n\neq j}=0$.)
We stress that, due to the limited control on the coefficients $I_{nj}$ (and on the $A^0-\eta_b$ interference),
the precise limits on this $(m_A,X_d)$ plane should be considered with caution, that is, as a qualitative trend, 
rather than a strict exclusion boundary. In view of the results of fig. \ref{fig:mixUps3S} and 
\ref{fig:mixUps1S} however, we conclude that values of $X_d$ beyond $\sim 2-3$ should lead to unacceptably 
large $BR(\Upsilon(nS)\to\gamma\,\tau^+\tau^-)$, at least within the range of the experimental constraints. 

Concerning the possibility of explaining the tension between the pQCD prediction and the observed $\eta_b(1S)$ 
mass, most of the ``favoured'' region, where this discrepancy can be interpreted (at least within $1\,\sigma$) 
by the presence of the $A^0$, is now excluded. The variations of $BR(\Upsilon(3S)\to\gamma\,\tau^+\tau^-)$ in 
the approximation of eq. (\ref{eq:Upsgameta2}), along this ``favoured region'', are shown in fig. 
\ref{fig:mixUps3SBR} (with $X_d$ now determined in terms of $m_A$ through eq. (\ref{eq:ma})). Only two mass 
ranges remain, where the tension between the observed and the pQCD-predicted $\eta_b^0$ masses may be 
interpreted as an $A^0$ effect: in the first case, $m_A\sim 9.39$~GeV, the required $X_d$ is sufficiently small 
to escape the {\sc BaBar} bounds. This coincides with the region of natural NMSSM couplings. Note however that 
the acceptable mass range is now only of a few $10$~MeV around $m_{\mbox{\tiny obs}}$, which supposes a very 
fine accidental relation between the $A^0$ and $\eta_b^0(1S)$ masses. The second region $m_A\gsim10.1$~GeV is 
beyond the reach of the {\sc BaBar} searches and requires large $X_d\sim20$.  We must therefore regard the 
possibility of generating the $\eta_b$ mass shift in the context of a light CP-odd Higgs only as a marginal 
possibility.
\vspace{.5cm}

To summarize, we reviewed experimental constraints from {\sc Cleo} and {\sc BaBar} on radiative $\Upsilon$ 
decays through a pseudoscalar. Even though the pure Higgs contribution is little controlled in the limit of 
low-energy photons, we showed that the mixing effect should lead to a sizable contribution, which can be used 
as a theoretical estimate for $BR(\Upsilon(3S)\to\gamma\,\tau^+\tau^-)$. Taking this estimate seriously, we 
obtain that the reduced coupling of the $A^0$ to $b$ quarks must be small, $X_d\lsim2$, even in the range 
$m_A\sim9-10.1$~GeV. Note however that this region with small couplings is that which is the most naturally 
achieved in the NMSSM (in {\em e.g.} the R or Peccei-Quinn symmetry limits). The possibility of interpreting 
the small tension between the observed and pQCD-predicted $\eta_b^0(1S)$ masses through a mixing with the $A^0$ 
is nevertheless largely narrowed and confined to the ranges $m_A\sim m_{\mbox{\tiny obs}}$ or $m_A\gsim10.1$~GeV.

\section*{Acknowledgments}
The author thanks U. Ellwanger, U. Nierste and C. Smith for useful discussions and comments. This work was 
supported by the EU Contract No. MRTN-CT-2006-035482, FLAVIAnet, and by DFG through project C6 of the 
collective research centre SFB-TR9.


\begin{thebibliography}{999}
\bibitem{Drees:1989du}
  M.~Drees and K.-I.~Hikasa,
  Phys.\ Rev.\  D {\bf 41} (1990) 1547.

\bibitem{SanchisLozano:2002pm}
  M.~A.~Sanchis-Lozano,
  Mod.\ Phys.\ Lett.\  A {\bf 17} (2002) 2265
  [arXiv:hep-ph/0206156];\\
  M.~A.~Sanchis-Lozano,
  Int.\ J.\ Mod.\ Phys.\ A \textbf{19} (2004) 2183 [arXiv:hep-ph/0307313];\\
  M.~A.~Sanchis-Lozano,
  PoS \textbf{HEP2005} (2006) 334 [arXiv:hep-ph/0510374].

\bibitem{McElrath:2005bp}
  B.~McElrath,
  Phys.\ Rev.\ D {\bf 72}, 103508 (2005) [arXiv:hep-ph/0506151].

\bibitem{Sanchis-Lozano:2006gx}
  M.~A.~Sanchis-Lozano,
  J.\ Phys.\ Soc.\ Jap.\  {\bf 76} (2007) 044101
  [arXiv:hep-ph/0610046].

\bibitem{Dermisek:2006py}
  R.~Dermisek, J.~F.~Gunion and B.~McElrath,
  Phys.\ Rev.\  D {\bf 76} (2007) 051105
  [arXiv:hep-ph/0612031].

\bibitem{Fullana:2007uq}
  E.~Fullana and M.~A.~Sanchis-Lozano,
  Phys.\ Lett.\  B {\bf 653} (2007) 67
  [arXiv:hep-ph/0702190].

\bibitem{Hodgkinson:2008ei}
  R.~N.~Hodgkinson, Phys.\ Lett.\  B {\bf 665} (2008) 219
  [arXiv:0802.3197 [hep-ph]].

\bibitem{Carena:2002bb}
  M.~Carena, J.~R.~Ellis, S.~Mrenna, A.~Pilaftsis and C.~E.~M.~Wagner,
  Nucl.\ Phys.\ B \textbf{659} (2003) 145 [arXiv:hep-ph/0211467].

\bibitem{Lee:2007ai}
  J.~S.~Lee and S.~Scopel,
  Phys.\ Rev.\  D {\bf 75} (2007) 075001
  [arXiv:hep-ph/0701221].

\bibitem{Han:2004yd}
  T.~Han, P.~Langacker and B.~McElrath,
  Phys.\ Rev.\ D {\bf 70} (2004) 115006 [arXiv:hep-ph/0405244].

\bibitem{Kraml:2006ga}
  S.~Kraml \textit{et al.},
  ``Workshop on CP studies and non-standard Higgs physics,''
  arXiv:hep-ph/0608079.

\bibitem{Ellwanger:2009dp}
  U.~Ellwanger, C.~Hugonie and A.~M.~Teixeira,
  arXiv:0910.1785 [hep-ph]. {\em (and ref. therein)}

\bibitem{dobr}
B.~A.~Dobrescu, G.~Landsberg and K.~T.~Matchev, Phys.\ Rev.\ D {\bf 63}
  (2001) 075003 [arXiv:\break hep-ph/0005308];\newline
B.~A.~Dobrescu and K.~T.~Matchev, JHEP {\bf 0009} (2000) 031 
  [arXiv:hep-ph/0008192].

\bibitem{hsearch1}
U.~Ellwanger, J.~F.~Gunion, C.~Hugonie and S.~Moretti,
arXiv:hep-ph/0305109 (in  ``Physics interplay of the LHC and the ILC'',
G.~Weiglein {\it et al.}  [LHC/LC Study Group], Phys.\ Rept.\  {\bf
426} (2006) 47); \\
U.~Ellwanger, J.~F.~Gunion, C.~Hugonie and S.~Moretti,
arXiv:hep-ph/0401228 (in ``The Higgs working group:
Summary report 2003'', K.~A.~Assamagan {\it et al.}  [Higgs Working
Group Collaboration],  arXiv:hep-ph/0406152).

\bibitem{dg1}  
R.~Dermisek and J.~F.~Gunion,
  Phys.\ Rev.\ Lett.\  {\bf 95} (2005) 041801 [arXiv:hep-ph/0502105].
  
\bibitem{Dermisek:2005gg}
  R.~Dermisek and J.~F.~Gunion, Phys.\ Rev.\ D \textbf{73} (2006) 111701
  [arXiv:hep-ph/0510322].
  
\bibitem{dg2}
  R.~Dermisek and J.~F.~Gunion, Phys.\ Rev.\  D {\bf 75} (2007) 075019
  [arXiv:hep-ph/0611142].
  
\bibitem{hsearch2}  
U.~Ellwanger, J.~F.~Gunion and C.~Hugonie, JHEP {\bf 0507} (2005) 041 
 [arXiv:hep-ph/0503203];\\
S.~Chang, P.~J.~Fox and N.~Weiner,
  JHEP {\bf 0608} (2006) 068 [arXiv:hep-ph/0511250];\\
P.~W.~Graham, A.~Pierce and J.~G.~Wacker, ``Four taus at the
  Tevatron,'' arXiv:hep-ph/0605162;\\
S.~Moretti, S.~Munir and P.~Poulose, Phys.\ Lett.\ B {\bf 644} (2007)
  241  [arXiv:hep-ph/0608233];\\
S.~Chang, P.~J.~Fox and N.~Weiner,
  Phys.\ Rev.\ Lett.\  {\bf 98} (2007) 111802 [arXiv:hep-ph/0608310];\\
T.~Stelzer, S.~Wiesenfeldt and S.~Willenbrock,
  Phys.\ Rev.\  D {\bf 75} (2007) 077701 [arXiv:hep-ph/0611242];\\
U.~Aglietti {\it et al.}, ``Tevatron-for-LHC report: Higgs,''
  arXiv:hep-ph/0612172;\\
 K.~Cheung, J.~Song and Q.~S.~Yan,
  Phys.\ Rev.\ Lett.\  {\bf 99} (2007) 031801 [arXiv:hep-ph/0703149].
  
\bibitem{Dermisek:2007yt}
  R.~Dermisek and J.~F.~Gunion,
  Phys.\ Rev.\  D {\bf 76} (2007) 095006
  [arXiv:0705.4387 [hep-ph]].

\bibitem{hsearch3}
M.~Carena, T.~Han, G.~Y.~Huang and C.~E.~M.~Wagner,  JHEP {\bf 0804}
  (2008) 092 [arXiv:0712.2466 [hep-ph]];\\ 
J.R. Forshaw, J.F. Gunion, L. Hodgkinson, A. Papaefstathiou and  A.D.
   Pilkington, JHEP {\bf 0804} (2008) 090 [arXiv:0712.3510 [hep-ph]];\\
A.~Djouadi {\it et al.},  JHEP {\bf 0807} (2008) 002
  [arXiv:0801.4321 [hep-ph]];\\
  S.~Chang, R.~Dermisek, J.~F.~Gunion and N.~Weiner,
  ``Nonstandard Higgs Boson Decays,''
  arXiv:0801.4554 [hep-ph];\\
A.~Belyaev {\it et al.}, ``The Scope of the 4 tau Channel in
  Higgs-strahlung and Vector Boson Fusion for the NMSSM No-Lose Theorem
  at the LHC,'' arXiv:0805.3505 [hep-ph].
  
\bibitem{ALEPH:2010aw}
  S.~Schael {\it et al.}  [ALEPH Collaboration],
  JHEP {\bf 1005} (2010) 049
  [arXiv:1003.0705 [hep-ex]].

\bibitem{Dermisek:2010mg}
  R.~Dermisek and J.~F.~Gunion,
  Phys.\ Rev.\  D {\bf 81}, 075003 (2010)
  [arXiv:1002.1971 [hep-ph]].

\bibitem{Upsilon}
  F.~Domingo, U.~Ellwanger, E.~Fullana, C.~Hugonie and M.~A.~Sanchis-Lozano,
  JHEP {\bf 0901} (2009) 061
  [arXiv:0810.4736 [hep-ph]]

\bibitem{Domingo:2009tb}
  F.~Domingo, U.~Ellwanger and M.~A.~Sanchis-Lozano,
  Phys.\ Rev.\ Lett.\  {\bf 103} (2009) 111802
  [arXiv:0907.0348 [hep-ph]].

\bibitem{CLEO:2008hs}
  W.~Love {\it et al.}  [CLEO Collaboration],
  Phys.\ Rev.\ Lett.\  {\bf 101} (2008) 151802
  [arXiv:0807.1427 [hep-ex]].

\bibitem{Aubert:2009cka}
  B.~Aubert {\it et al.}  [BABAR Collaboration],
  Phys.\ Rev.\ Lett.\  {\bf 103} (2009) 181801
  [arXiv:0906.2219 [hep-ex]].

  B.~Aubert {\it et al.}  [BABAR Collaboration],
  Phys.\ Rev.\ Lett.\  {\bf 103} (2009) 081803
  [arXiv:0905.4539 [hep-ex]].

\bibitem{Guido:2009zz}
  E.~Guido  [BABAR Collaboration],
  PoS E {\bf PS-HEP2009} (2009) 375.


\bibitem{eta_b:pqcd}
  S.~Recksiegel and Y.~Sumino,
  Phys.\ Lett.\  B {\bf 578} (2004) 369
  [arXiv:hep-ph/0305178].

  B.~A.~Kniehl, A.~A.~Penin, A.~Pineda, V.~A.~Smirnov and M.~Steinhauser,
  Phys.\ Rev.\ Lett.\  {\bf 92} (2004) 242001
  [arXiv:hep-ph/0312086]. 

  A.~A.~Penin,
  ``The mass of $\eta_b$,''
  arXiv:0905.4296 [hep-ph].

\bibitem{Brambilla}
  N.~Brambilla {\it et al.}  [Quarkonium Working Group],
  ``Heavy quarkonium physics,''
  arXiv:hep-ph/0412158.

  S.~Meinel,
  arXiv:1007.3966 [hep-lat].

\bibitem{Gorbahn:2009pp}
  M.~Gorbahn, S.~Jager, U.~Nierste and S.~Trine,
  arXiv:0901.2065 [hep-ph].

\bibitem{NMSSMTools}
  U.~Ellwanger, J.~F.~Gunion and C.~Hugonie,
  JHEP {\bf 0502} (2005) 066
  [arXiv:hep-ph/0406215].

  U.~Ellwanger and C.~Hugonie,
  Comput.\ Phys.\ Commun.\  {\bf 175} (2006) 290
  [arXiv:hep-ph/0508022].

http://www.th.u-psud.fr/NMHDECAY/nmssmtools.html

\bibitem{Andreas:2010ms}
  S.~Andreas, O.~Lebedev, S.~Ramos-Sanchez and A.~Ringwald,
  arXiv:1005.3978 [hep-ph].

\bibitem{BNMSSM}
  F.~Domingo and U.~Ellwanger,
  JHEP {\bf 0712}, 090 (2007)
  [arXiv:0710.3714 [hep-ph]].

\bibitem{Wilczek:1977pj}
  F.~Wilczek,
  Phys.\ Rev.\ Lett.\  \textbf{40} (1978) 279.

 H.~E.~Haber, A.~S.~Schwarz and A.~E.~Snyder,
  Nucl.\ Phys.\  B {\bf 294} (1987) 301.

\bibitem{guide}
  J.~F.~Gunion, H.~E.~Haber, G.~L.~Kane and S.~Dawson,
  {\it The Higgs Hunter's Guide} (Perseus Publishing, Cambridge (US), 
  MA, 1990).

\bibitem{PDG} C. Amsler et al. (Particle Data Group), Physics Letters B667, 1 (2008) and
2009 partial update for the 2010 edition 

\bibitem{BABAR:2008vj}
  B.~Aubert {\it et al.}  [BABAR Collaboration],
  Phys.\ Rev.\ Lett.\  {\bf 101} (2008) 071801
  [Erratum-ibid.\  {\bf 102} (2009) 029901]
  [arXiv:0807.1086 [hep-ex]].

\bibitem{BABAR:2009pz}
[BABAR Collaboration],
  Phys.\ Rev.\ Lett.\  {\bf 103} (2009) 161801
  [arXiv:0903.1124 [hep-ex]].

\bibitem{Godfrey:2001eb}
  S.~Godfrey and J.~L.~Rosner,
  Phys.\ Rev.\  D {\bf 64} (2001) 074011
  [Erratum-ibid.\  D {\bf 65} (2002) 039901]
  [arXiv:hep-ph/0104253].

\bibitem{oliver} A.~Le~Yaouanc et al., {\it Hadron transitions in the
quark model,} (Gordon and Breach Science Publishers, 1988).

\bibitem{Rashed:2010jp}
  A.~Rashed, M.~Duraisamy and A.~Datta,
  arXiv:1004.5419 [hep-ph].

\end{thebibliography}
\end{document}